\documentclass[letterpaper]{article}
\usepackage{aaai16}
\usepackage{times}
\usepackage{helvet}
\usepackage{courier}
\usepackage{fixbib}
\usepackage{url}
\usepackage{amsmath}
\usepackage{amssymb}
\usepackage{color}
\usepackage{epsf}
\usepackage{color}
\usepackage{graphicx}
\usepackage{epsfig}
\usepackage{fancybox}
\usepackage{amssymb}
\usepackage{amsmath}
\usepackage{multicol}
\usepackage{hhline}
\usepackage{xspace,epic,eepic,graphicx}
\usepackage{latexsym}
\usepackage[all]{xy}

\mathchardef\myhyphen="2D

\newcommand{\boxtheorem}{\hfill $\blacksquare$\vspace{1mm}}
\newcommand{\ignore}[1]{}
\newcommand{\nit}[1]{{\it #1}}
\newcommand{\Q}{\mathcal{Q}}

\DeclareMathAlphabet{\mathpzc}{OT1}{pzc}{m}{it}
\newcommand{\match}{{\mathpzc m}}
\newcommand{\lub}{{\it lub}}
\newcommand{\glb}{{\it glb}}

\newcommand{\eat}[1]{}
\newcommand{\mc}[1]{\mathcal{ #1}}

\newcommand{\n}{~{\it not}~}

\newcommand{\ma}{M_{\!A}}

\newcommand{\conp}{$\nit{co}\mbox{-}\!\nit{NP}$}

{\vskip \abovedisplayskip \refstepcounter{lemmaA-counter}%
\noindent {\bf Lemma A.\arabic{lemmaA-counter}.}}%

{\vskip \abovedisplayskip \refstepcounter{definitionA-counter}%
\noindent {\bf Definition A.\arabic{definitionA-counter}.}}%

\newcommand{\comlb}[1]{{\vspace{2mm}\noindent \bf \underline{COMM(LEO):}}~ #1 \hfill {\bf END.}\\}

\newcounter{theorem-counter}
\newcounter{corollary-counter}
\newcounter{lemma-counter}
\newcounter{definition-counter}
\newcounter{example-counter}
\newcounter{proposition-counter}
\newcounter{remark-counter}
\newcounter{definitionA-counter}
\newcounter{lemmaA-counter}
\newcounter{propositionA-counter}
{\vskip \abovedisplayskip \refstepcounter{theorem-counter}%
\noindent {\bf Theorem \arabic{theorem-counter}.}}%
{\vskip \abovedisplayskip \refstepcounter{corollary-counter}%
\noindent {\bf Corollary \arabic{corollary-counter}.}}%
{\vskip \abovedisplayskip \refstepcounter{lemma-counter}%
\noindent {\bf Lemma \arabic{lemma-counter}.}}%

\newenvironment{definition}%
{\vskip \abovedisplayskip \refstepcounter{definition-counter}%
\noindent {\bf Definition \arabic{definition-counter}.}}%
\newenvironment{example}%
{\vskip \abovedisplayskip \refstepcounter{example-counter}%
\noindent {\bf Example \arabic{example-counter}.}}%
{\vskip \abovedisplayskip \refstepcounter{proposition-counter}%
\noindent {\bf Proposition \arabic{proposition-counter}.}}%
{\vskip \abovedisplayskip \refstepcounter{remark-counter}%
\noindent {\bf Remark \arabic{remark-counter}.}}%
\title{{\bf Enforcing Relational Matching Dependencies with Datalog  for Entity Resolution}}

\author{{\bf \large Zeinab Bahmani} \and {\bf \large Leopoldo Bertossi}\thanks{Contact author. Research funded by NSERC Discovery.}\\
Carleton University, Ottawa, Canada. \\
zbahmani@connect.carleton.ca, \ \  bertossi@scs.carleton.ca}

\begin{document}
%\sloppy
\maketitle

\pagestyle{plain} %DELETE THIS LINE AT THE END

\begin{abstract}
Entity resolution (ER)  is about
identifying and merging records in a database that represent the same \eat{external} real-world
entity. Matching dependencies (MDs) have been introduced and investigated as declarative
rules that specify ER policies. An ER process induced by MDs over a dirty instance leads to multiple
clean instances, in general. General {\em answer sets programs} have been proposed to specify the MD-based cleaning
task and its results. In this work, we extend MDs to {\em relational MDs}, which capture more application semantics, and identify
classes of relational MDs for which the general ASP can be automatically rewritten into a stratified Datalog program, with the single
clean instance as its standard
model.
\end{abstract} \vspace{-3mm}

%%%%%%%%%%%%%%%%%%%%%%%%%%%%%%%%%%%%%%%%%%%%%%%%%%%%%%%
\section{Introduction}

The presence in a database of duplicate, but non-identical representations of the same external entity  leads to
uncertainty. Applications running on top of the database or a query answering process may not be able to tell them apart, and the
results may lead to ambiguity, semantic problems, such as unintended inconsistencies, and erroneous results. In this situation, the database
has to be cleaned. The whole area of {\em entity resolution} (ER) deals with identifying and
merging database records in a database that refer to the same real-world
entity \cite{naumannACMCS,elmargamid}. In so doing, duplicates are
  eliminated from the database, while at the same time new tuples are created through the merging process. ER is one of the most common and difficult problems in data cleaning.

In the last few years there has been strong and increasing interest in providing declarative and generic solutions to data cleaning problems \cite{BertossiBravo}, in particular,
in logical specifications of the ER process. In this direction, {\em matching dependencies} (MDs) have been proposed \cite{Fan08,FanJLM09}. They are declarative rules that  assert
 that certain attribute values in relational tuples have to be merged, i.e. made identical,
when certain similarity conditions hold between possibly other attribute values in those tuples.

  \begin{example} \label{ex:first} Consider the relational predicate $R(A,B)$,  with attributes $A$ and $B$. The symbolic rule \ $R[A] \approx R[A]  \rightarrow  R[B] \doteq R[B]$ \
is an MD specifying that, if for any two database tuples $R(a_1,b_1), R(a_2,b_2)$ in an instance $D$, when $A$-values are similar, i.e.\ $a_1 \approx a_2$, then their $B$-values
have to be made equal (merged), i.e.\ $b_1$ or $b_2$ (or both) have to be changed to a value in
 common.

 Let us assume that $\approx$ is reflexive
and symmetric, and that $a_2 \approx a_3$, but $a_2 \not \approx a_1 \not \approx a_3$.
The table on the left-hand side (LHS) below provides  the extension
for predicate $R$ in $D$. In it some duplicates are not ``resolved",  e.g.\
the tuples (with tuple identifiers) $t_1$ and
 $t_2$ have similar -- actually equal -- $A$-values, but their $B$-values are different.

{\small
\begin{center}
\begin{tabular}{c|c|c|}\hline
$R(D)$&$A$ & $B$  \\ \hline
$t_1$&$a_1$ & $b_1$  \\
$t_2$&$a_1$ & $b_2$  \\
$t_3$&$a_2$ & $b_3$  \\
$t_4$&$a_3$ & $b_4$ \\ \cline{2-3}
\end{tabular}
~~~~~~~~~~~~\begin{tabular}{c|c|c|}\hline
$R(D')$&$A$ & $B$  \\ \hline
$t_1$&$a_1$ & $b_1$  \\
$t_2$&$a_1$ & $b_1$  \\
$t_3$&$a_2$ & $b_5$  \\
$t_4$&$a_3$ & $b_5$ \\ \cline{2-3}
\end{tabular}
\end{center} }

\noindent $D$ does not satisfy the MD, and is a \emph{dirty} instance. After applying the
 MD, we could get the instance $D'$ on the right-hand side (RHS), where values for $B$ have been identified. %In principle, nothing prevents us from choosing a new value $b_5$ from the domain to do the matching.
$D'$ is \emph{stable} in the sense that the MD holds in the traditional sense
of an implication and ``$=$" on $D'$, which we call a \emph{clean instance}. In general, for a dirty instance
and a set of MDs, multiple clean instances may exist. Notice that if we add the MD $R[B] \approx R[B]  \rightarrow  \ R[A] \doteq R[A]$, creating a set of {\em interacting} MDs, a merging with one MD may create new similarities that
enable the other MD.
 \boxtheorem
\end{example}
A {\em dynamic semantics} for MDs was introduced in \cite{FanJLM09},  that requires pairs of
instances: a first one where the similarities hold, and a second where the mergings are enforced, e.g. $D$ and $D'$ in Example \ref{ex:first}. MDs, as introduced in \cite{FanJLM09}, do not specify what values to use when
merging two attribute values.

The semantics
was  refined and extended in  \cite{tocs} by means of {\em matching functions} (MFs) providing values  for  equality enforcements.  An
MF induces a lattice-theoretic structure on an attribute's domain.
Actually,  a {\em chase-based} semantics for MD enforcement was proposed. On this basis,
given an instance $D$ and a set $\Sigma$ of MDs, wrt. which $D$ may contain duplicates, the {\em chase procedure} may lead to
several different {\em clean and stable solutions} $D'$. Each of them can be obtained  by means of
a provably terminating, but non-deterministic, iterative procedure that
enforces the MDs through  application of MFs. The set of all such clean instances is denoted by $\mc{C}(D,\Sigma)$. Each clean instance can be seen as the result of an {\em uncertainty reduction} process.
If at the end there are several possible clean instances, uncertainty is still present, and expressed through this class of {\em possible worlds}. Identifying cases for which a single clean instance exists is particularly relevant:  for them uncertainty can be eliminated.

In \cite{kr},  a declarative specification of this procedural
data cleaning semantics was proposed. More precisely, a general methodology was developed to produce, from $D$, $\Sigma$ and the MFs,  an {\em answer set program} (ASP) \cite{gelfond91,brewka11} whose models are exactly the clean instances in the class $\mc{C}(D,\Sigma)$. The ASP enables reasoning in the presence of uncertainty due to  multiple clean instances. Computational implementations of ASP can be then used for reasoning, for computing clean instances, and for computing {\em certain query answers} (aka. {\em clean answers}), i.e. those that hold in all the clean instances \cite{kr}. Disjunctive ASPs, aka.  {\em disjunctive Datalog programs with stable model semantics} \cite{eiterGottlob97}, are used (and provably required) for this task.

For some classes of MDs, for any given initial instance $D$, the class $\mc{C}(D,\Sigma)$ contains a  {\em single clean instance} that can be computed in polynomial time in the size of $D$. Some sufficient syntactic and MF-dependent conditions were identified in \cite{tocs}.
In this work we identify a new important ``semantic" class of MDs, where the initial instance is also considered. This is the  {\em similarity-free attribute intersection class} (the {\em SFAI} class) of {\em combinations of MDs and initial instances}. Members of this class also have {\em (polynomial-time computable) single clean instances}. For all these classes, we show that the general
ASP mentioned above can be automatically and syntactically transformed into an equivalent {\em stratified Datalog} program with the single clean instance as its {\em standard model}, which can be computed bottom-up from $D$ in polynomial time in the size of $D$ \cite{ahv95,CGT89}.

Relational ER has been approached by the machine learning community \cite{lise}. The idea is to learn from examples a classifier that can be used to determine if an arbitrary pair of records (or tuples), $r_1, r_2$, are duplicates
(or each other) or not. In order to speed up the process of learning and applying the classifier,  usually {\em blocking} techniques are applied \cite{blocking}. They are used to group records in clusters (blocks), for further comparison of pairs within clusters, but never of two records in different clusters. Interestingly, as reported in \cite{sum15}, MDs can be used in the blocking phase. As expected, MDs were also used during
the final merging phase, after the calls to the classifier. However, the use at the earlier stage is rather surprising. The kind of MDs in this case turn out to belong, together with the initial instance,  to the SFAI class.
Actually, this allowed  implementation of MD-based blocking by means of Datalog.

The reason for using MDs at the blocking stage is that they may convey semantic relationships between records for different entities, and can then be used to {\em collectively} block records for different entities \cite{lise}: blocking together
two records for an entity, say of books,  may depend on having blocked together related records for a different entity, say of authors.
For these kinds of applications, to capture semantic relationships, MDs were extended with relational atoms (conditions) in the antecedents, leading to the class of {\em relational MDs}.

In this work  we also introduce and investigate the class of relational MDs, we extend the single-clean instance classes mentioned above to the relational MD case, and we obtain in a uniform manner
Datalog programs for the enforcement of MDs in these classes. For lack of space, our presentation is based mainly on representative examples.

\ignore{
This paper is structured as follows. Section \ref{sec:prel} introduces the necessary background on relational databases, matching dependencies and their semantics, and disjunctive Datalog programs. In Section
\ref{sec:decl}, we introduce the cleaning programs that specify the clean instances wrt.  a set of MDs. They are used
in Section \ref{sec:QA} to do clean query answering. In Section \ref{sec:analysis} we analyze and transform the cleaning programs, addressing some complexity issues. In Section \ref{sec:swoosh} we present a declarative version
of the union case of Swoosh. In Section \ref{sec:concl}, we obtain a few final conclusions.
Many more details in general, full proofs of results, and also examples with the {\em DLV} system \cite{dlv}, can be found in the extended version of this paper \cite{zeinab11}. }

\vspace{-2mm}
\section{Background} \label{sec:prel}

\vspace{-1mm}
We consider relational schemas $\mc{R}$ with a possibly infinite data domain $U$, a finite set of
database predicates, e.g.\ $R$, and a set of built-in predicates, e.g.\ $=, \neq$. Each $R \in \mc{R}$ has attributes, say $A_1, \ldots, A_n$, each of them
with a domain $\nit{Dom}_{\!A_i} \subseteq U$.  We may assume that the $A_i$s are different, and different predicates
do not share attributes. However, different attributes may share the same domain.

An instance $D$ for $\mc{R}$ is a finite set of ground atoms (or tuples) of the form $R(c_1,\ldots, c_n)$, with $R \in \mc{R}$,
$c_i \in \nit{Dom}_{\!A_i}$.  We will assume that tuples
have identifiers, as in Example \ref{ex:first}. They allow us to compare extensions of the same
predicate in different instances, and trace changes of attribute values.
Tuple identifiers can be accommodated by adding to each predicate $R \in \mc{R}$ an extra attribute, $T$, that acts as a key. Then, tuples take the form $R(t,c_1,\ldots, c_n)$, with $t$ a value for $T$.
 Most of the
time we leave the tuple identifier implicit, or we use it to denote the whole tuple. More precisely, if $t$ is a tuple identifier in an instance $D$, then $t^{D}$ denotes the entire atom, $R(\bar{c})$, identified by $t$.
Similarly, if $\cal{A}$ is a list of attributes of predicate $R$, then $t^D[\mc{A}]$ denotes the tuple identified by $t$, but restricted to the attributes in $\cal{A}$. We assume that tuple identifiers are unique across the entire instance.

\ignore{
Schema $\mc{R}$ determines a language $L(\mc{R})$ of first-order (FO) predicate logic. A conjunctive query is a formula of
$L(\mc{R})$ of the form \ $\mc{Q}(\bar{x})\!: \ \exists \bar{y}(P_1(\bar{x}_1) \wedge \cdots \wedge P_m(\bar{x}_m))$,
where $P_i \in \mc{R}$, $\bar{x} = (\cup_i \bar{x}_i) \smallsetminus \bar{y}$ is the list free variables
of the query, say $\bar{x} = x_1\cdots x_k$. An answer to $\mc{Q}$ in instance $D$ is a sequence $\langle a_1, \ldots, a_k\rangle
\in U^k$ that makes $\mc{Q}$ true in $D$, denoted $D \models \mc{Q}[a_1,\ldots,a_k]$. $\mc{Q}(D)$ denotes the set of answers to $\mc{Q}$ in $D$, and can be seen and treated as an instance for an ``answer" relational schema, possibly different from the original one.  }

For a schema $\mc{R}$ with predicates $R_1[\bar{L}_1],$ $R_2[\bar{L}_2]$, with lists of attributes
$\bar{L}_1, \bar{L}_2$, resp.,
a {\em matching dependency} (MD) \cite{FanJLM09} is an expression of the form:
\begin{equation}
\varphi\!:  \ \ R_1[\bar{X}_1] \approx R_2[\bar{X}_2] \ \longrightarrow \ R_1[\bar{Y}_1] \doteq R_2[\bar{Y}_2]. \label{eq:md2} \vspace{-4.5mm}
\end{equation}

\phantom{ooo \linebreak}
\noindent Here, $\bar{X}_1, \bar{Y}_1$ are sublists of $\bar{L}_1$, and $\bar{X}_2,\bar{Y}_2$ sublists of $\bar{L}_2$. The lists $\bar{X}_1, \bar{X}_2$ (also $\bar{Y}_1, \bar{Y}_2$) are
{\em comparable}, i.e.\  the attributes in them, say $X_1^j, X_2^j$, are {\em pairwise comparable} in the sense that they share
 the same data domain $\nit{Dom}_j$ on which a binary similarity (i.e.\  reflexive and symmetric) relation $\approx_j$ is defined. \ignore{Actually, (\ref{eq:md2}) can be seen as an abbreviation for
\vspace{1mm}
\centerline{$\varphi\!:  \ \ \bigwedge R_1[X_1^j] \approx_j R_2[X_2^j] \ \longrightarrow \ \bigwedge R_1[Y_1^k] \doteq R_2[Y_2^k]. $}
\vspace{1mm}
\noindent }

 The MD (\ref{eq:md2}) intuitively states that if, for an $R_1$-tuple $t_1$ and an $R_2$-tuple $t_2$
in an instance $D$ the attribute values in $t_1^D[\bar{X}_1]$ are similar to attribute
values in $t_2^D[\bar{X}_2]$, then the values $t_1^D[\bar{Y}_1]$ and $t_2^D[\bar{Y}_2]$
have to be made identical. This update results in another instance $D'$, where  $t_1^{D'}[\bar{Y}_1] = t_2^{D'}[\bar{Y}_2]$ holds.
W.l.o.g., we may assume that the list of attributes on the RHS of
MDs contain only one conjunct (attribute).

For a set $\Sigma$ of MDs, a pair of instances $(D,D')$ satisfies $\Sigma$
if whenever $D$ satisfies the antecedents of the MDs, then  $D'$ satisfies the consequents
(taken as equalities).
If $(D,D) \not \models \Sigma$, we say that $D$ is ``dirty" (wrt.  $\Sigma$). On the other hand,
an instance $D$ is {\em stable} if $(D,D) \models \Sigma$ \ \cite{FanJLM09}.

We now review some elements in \cite{tocs}. In order to {\em enforce} an MD on two tuples, making values of attributes identical,
we assume that for each comparable pair of attributes $A_1,A_2$
with domain (in common) ${\it Dom}_{\!A}$, there is a binary {\em matching function} (MF)
$\match_A: {\it Dom}_{\!A}\times {\it Dom}_{\!A}\rightarrow {\it Dom}_{\!A}$, such that
$\match_A(a,a')$ is used to replace two values $a, a' \in {\it Dom}_{\!A}$ whenever necessary.
MFs are idempotent, commutative, and associative. \ignore{
The structure $({\it Dom}_{\!A}, \match_A)$  forms a {\em join semilattice}, that is,
a partial order with a  least upper bound ($\lub$) for every pair of elements. The induced
partial order $\preceq_A$ on ${\it Dom}_{\!A}$ is defined by: \ $a \preceq_A a'$ whenever $\match_A(a,a') = a'$. }
Similarity relations
 and MFs are treated as built-in  relations.

A chase-based semantics for entity resolution with MDs is as follows:
 starting from an instance $D_0$, we identify pairs of tuples $t_1,t_2$
that satisfy the similarity conditions on the left-hand side of an MD  $\varphi$,
i.e.\  $t_1^{D_0}[\bar{X}_1] \approx t_2^{D_0}[\bar{X}_2]$ (but not the identity in its RHS),
 and apply an MF on the
values for the right-hand side attribute,  $t_1^{D_0}[A_1],t_2^{D_0}[A_2]$, to make them both
equal to $\match_A(t_1^{D_0}[A_1],t_2^{D_0}[A_2])$. We keep doing this on the resulting instance,
in a {\em chase-like} procedure \cite{ahv95}, until a stable instance is reached (cf.
\cite{tocs} for details), i.e. a {\em clean instance}.
\ignore{
\begin{definition}
\label{def:immediate}  Let $D,D'$ be database instances with the same set of tuple identifiers,
$\Sigma$ be a set of MDs, and
$\varphi: R_1[\bar{X}_1] \approx R_2[\bar{X}_2] \rightarrow R_1[\bar{Y}_1] \doteq R_2[\bar{Y}_2]$ be
an MD in $\Sigma$. Let
$t_1,t_2$ be an $R_1$-tuple and an $R_2$-tuple identifiers, respectively, in both $D$ and $D'$.
Instance $D'$ is the {\em immediate result of enforcing} $\varphi$ on $t_1,t_2$ in
instance $D$, denoted $(D,D')_{[t_1,t_2]} \models \varphi$, if
\begin{itemize}
\item [(a)] $t_1^D[\bar{X}_1] \approx t_2^D[\bar{X}_2]$, but $t_1^D[\bar{Y}_1] \neq t_2^D[\bar{Y}_2]$;
\item [(b)] $t_1^{D'}[\bar{Y}_1] = t_2^{D'}[\bar{Y}_2] = \match_A(t_1^D[\bar{Y}_1], t_2^D[\bar{Y}_2])$; and
\item [(c)] $D,D'$ take the same values on every other tuple and attribute. \boxtheorem
\end{itemize}
\end{definition}

\vspace{-3mm}
\begin{definition}
\label{clean-def}
For an instance $D_0$ and a set $\Sigma$ of MDs , an instance $D_k$ is
{\em $(D_0,\Sigma)$-clean} if $D_k$ is stable,
and there exists a finite sequence of
instances $D_1,\ldots, D_{k-1}$ such that, for every $i \in [1,k]$,
$(D_{i-1},D_i)_{[t_1^i,t_2^i]} \models \varphi$, for some $\varphi \in \Sigma$ and tuple
identifiers $t_1^i,t_2^i$.\boxtheorem
\end{definition}  }
An instance $D_0$ may have several $(D_0,\Sigma)$-clean instances.
 $\mc{C}(D_0,\Sigma)$ denotes the set of clean instances for $D_0$
wrt.\ $\Sigma$.

For given $D$ and $\Sigma$, the class of clean instances can be specified as the stable models of a  logic program $\Pi(D_0,\Sigma)$ in $\nit{Datalog}^{\!\!\vee,\!\!\n}$, i.e.\   a disjunctive Datalog program with weak negation and stable model semantics \cite{gelfond91,eiterGottlob97}, with rules of the form:
\ $A_1 \vee \ldots \vee A_n \leftarrow P_1, \ldots, P_m, \n N_1, \ldots, \n N_k$. Here,
  $0\leq n,m,k$, and $A_i, P_j, N_s$ are (positive) atoms. Rules with $n=0$ are called {\em program constraints} and have the effect of eliminating
  the stable models of the program (without them) that make their bodies (RHS of the arrow) true. When $n=1$ and $k=0$, we have (plain) {\em Datalog} programs. When $n\geq 1$ and
  \nit{not} is stratified, we have {\em disjunctive, stratified Datalog} programs, denoted $\nit{Datalog}^{\!\!\vee,\!\!\n\!\!,s}$. The subclass with $n=1$ is {\em stratified Datalog},
  denoted $\nit{Datalog}^{\!\!\n\!\!,s}$\!\!.

We  now introduce general cleaning programs by means of a representative example (for full generality and details, see
\cite{kr}).
Let $D_0$ be a given, possibly dirty initial instance wrt.  a set $\Sigma$ of MDs. The {\em cleaning program}, $\Pi(D_0,\Sigma)$, that we will
 introduce here, contains an $(n+1)$-ary predicate $R'_i$, for each $n$-ary database predicate $R_i$. It will be used
 in the form $R'_i(T,\bar{Z})$, where $T$ is a
 variable for  the tuple identifier attribute, and $\bar{Z}$ is a list
of variables standing for the (ordinary) attribute values of $R_i$.

For every attribute $A$ in the schema, with domain  ${\it Dom}_{\!A}$, the built-in ternary predicate
$M_{\!A}$ represents the MF $\match_A$, i.e.\  $\ma(a,a',a'')$ means
$\match_A(a,a') = a''$.  $X$ $\preceq_A$ $Y$ is used as an abbreviation for
$\ma(X,Y,Y)$. For attributes $A$ without a matching function,
 $\preceq_A$ becomes the equality, $=_A$. For lists of variables $\bar{Z}_1= \langle Z^1_1,\ldots Z^n_1 \rangle$
and $\bar{Z}_2= \langle Z^1_2,\ldots Z^n_2 \rangle$,  $\bar{Z}_1 \preceq \bar{Z}_2$
denotes the conjunction $Z^1_1 \preceq_{A_1} Z^1_2 \land \ldots \land Z^n_1 \preceq_{A_n} Z^n_2$.
Moreover, for each attribute $A$, there is a built-in binary predicate $\approx_A$.
 For two lists of variables $\bar{X}_1=\langle X_1^1,\ldots X_1^l \rangle$
and  $\bar{X}_2=\langle X_2^1,\ldots X_2^l \rangle$ representing comparable attribute
values, $\bar{X}_1 \approx \bar{X}_2$ denotes  the conjunction
$X_1^1 \approx_1 X_2^1 \land \ldots \land X_1^l \approx_l X_2^l$.

In intuitive terms, program $\Pi(D_0,\Sigma)$ has rules to {\em implicitly simulate} a chase sequence, i.e.\  rules that
enforce MDs on pairs of tuples that satisfy certain similarities, create newer versions of those tuples
by applying matching functions, and make the older versions of the tuples unavailable for other rules.
The main idea is making stable models of the program correspond to valid chase sequences leading to clean instances.

When the conditions for applying an MD hold, we have the choice between matching or not.\footnote{Matching is merging, or making identical, two attribute values on the basis of the MDs.} If we do, the
tuples are updated to new versions. Old versions are collected in a predicate, and tuples that have not participated
in a matching  that was possible never become old versions (see the last denial constraint under 2. in Example \ref{ex:first1}, saying that the RHS of the arrow cannot be made true).

The program eliminates, using {\em program  constraints},
instances (models of the program) that are the result of an {\em illegal} set of applications of MDs, i.e.
they cannot put them in a linear (chronological) order
representing chase steps. This occurs when matchings use old versions
of tuples that have been replaced by new versions.
To ensure that the matchings are enforced according to an order that
correctly represents a
chase,  pairs of matchings are stored in an auxiliary relation,  ${\it
Prec}$. The last two program constraints under 6. in the example  make ${\it Prec}$ a linear order. In particular, matchings  performed using old versions of tuples are disallowed.

\begin{example} \label{ex:first1} Consider relation $R(A,B)$ with extension in $D_0$ as below; and assume that
exactly
the following similarities hold: $a_1\approx a_2$, $b_2\approx b_3$; and the MFs are as follows: \vspace{-4mm}
{\small \begin{multicols}{2}
$M_B(b_1, b_2,b_{12})$,

 $M_B(b_2, b_3,b_{23})$,

   $M_B(b_1, b_{23},b_{123})$,

  $M_B(b_3, b_4,b_{34})$.
\begin{center}
\begin{tabular}{c|c|c|}\hline
$R(D_0)$&$A$ & $B$ \\ \hline
$t_1$&$a_1$ & $b_1$ \\
$t_2$&$a_2$ & $b_2$ \\
$t_3$&$a_3$ & $b_3$ \\ \cline{2-3}
\end{tabular}
\end{center}
\end{multicols} }

\vspace{-2mm} \noindent $\Sigma$ contains the MDs:
{\small \begin{center}
$\varphi_1: R\left[A\right] \approx R\left[A\right] \rightarrow R\left[B\right] \doteq R\left[B\right]$,\\
$\varphi_2: R\left[B\right] \approx R\left[B\right] \rightarrow R\left[B\right] \doteq R\left[B\right]$,
\end{center}}
\noindent which are  {\em interacting} in that the set of attributes in the RHS of $\varphi_1$, namely $\{R[B]\}$, and the set of attributes in the LHS of $\varphi_2$, namely $\{R[B]\}$, have non-empty intersection. For the same reason, $\varphi_2$ also interacts with itself.
Enforcing $\Sigma$ on $D_0$  results in two alternative chase
sequences, each enforcing the MDs in a different order, and two final stable clean instances
$D_1$ and $D'_2$.

\vspace{1mm}
{\footnotesize
\hspace*{-2mm}
\begin{tabular}{c|c|c|}\hline
$D_0$&$A$ & $B$ \\ \hline
$t_1$&$a_1$ & $b_1$ \\
$t_2$&$a_2$ & $b_2$ \\
$t_3$&$a_3$ & $b_3$ \\ \cline{2-3}
\end{tabular} $\Rightarrow_{\varphi_1}$ \hspace*{-2mm}
\begin{tabular}{c|c|c|}\hline
$D_1$&$A$ & $B$ \\ \hline
$t_1$&$a_1$ & $b_{12}$ \\
$t_2$&$a_2$ & $b_{12}$ \\
$t_3$&$a_3$ & $b_3$ \\ \cline{2-3}
\end{tabular} \\ \ \\ \
\hspace*{-2mm}
\begin{tabular}{c|c|c|}\hline
$D_0$&$A$ & $B$ \\ \hline
$t_1$&$a_1$ & $b_1$ \\
$t_2$&$a_2$ & $b_2$ \\
$t_3$&$a_3$ & $b_3$ \\ \cline{2-3}
\end{tabular} $\Rightarrow_{\varphi_2}$ \hspace*{-2mm}
\begin{tabular}{c|c|c|}\hline
$D'_1$&$A$ & $B$ \\ \hline
$t_1$&$a_1$ & $b_1$ \\
$t_2$&$a_2$ & $b_{23}$ \\
$t_3$&$a_3$ & $b_{23}$ \\ \cline{2-3}
\end{tabular} $\Rightarrow_{\varphi_1}$ \hspace*{-2mm}
\begin{tabular}{c|c|c|}\hline
$D'_2$&$A$ & $B$ \\ \hline
$t_1$&$a_1$ & $b_{123}$ \\
$t_2$&$a_2$ & $b_{123}$ \\
$t_3$&$a_3$ & $b_{23}$ \\ \cline{2-3}
\end{tabular}
}
\vspace{2mm}\noindent The cleaning program $\Pi(D_0,\Sigma)$ is as follows:

\vspace{2mm}
{\small \begin{enumerate}
\item
\ignore{$\nit{Dom}(a_i)$. ~~~~~~~~~~~$\nit{Dom}(b_i)$.~~~~~~~~~~~~($1 \leq i \leq 3$)\\
~$\nit{Dom}(b_{12})$. ~~~~~~~~~$\nit{Dom}(b_{23})$. ~~~~~~~~~$\nit{Dom}(b_{123})$.\\}
 $R'(t_1, a_1, b_1).$  $R'(t_2, a_2, b_2).$  $R'(t_3, a_3, b_3).$  \ {\scriptsize (plus $M_B$ facts)}
\ignore{$M_{\!B}\!(b_1, b_2,b_{12}).$ \ $M_{\!B}\!(b_2, b_3,b_{23}).$ \ $M_{\!B}\!(b_1, b_{23},b_{123}).$} \vspace{-6mm}
\end{enumerate} }
{\small
\begin{eqnarray*}
~~~2.~{\it Match}_{\!\varphi_{\!1}}\!(T_1, X_1, Y_1, T_2, X_2, Y_2)\; \vee\hspace{2.5cm}\\
{\it NotMatch}_{\!\varphi_{\!1}}\!(T_1, X_1, Y_1, T_2, X_2, Y_2)\;\; \leftarrow \hspace{1.3cm}\\
R'(T_1, X_1, Y_1),\; R'(T_2, X_2, Y_2),\;X_1 \approx X_2, \;  Y_1 \neq Y_2.\hspace{-7mm}&\\
{\it Match}_{\!\varphi_{\!2}}\!(T_1, X_1, Y_1, T_2, X_2, Y_2)\; \vee\hspace{2.5cm}\\
{\it NotMatch}_{\!\varphi_{\!2}}\!(T_1, X_1, Y_1, T_2, X_2, Y_2)\;\; \leftarrow \hspace{1.3cm}\\
~~~~~~~~~~~R'(T_1, X_1, Y_1),\; R'(T_2, X_2, Y_2),\;Y_1 \approx Y_2, \;  Y_1 \neq Y_2.\hspace{-5mm}
\end{eqnarray*}}
\vspace{-6mm}{\small
\begin{align*}
&~~~~~~~{\it Match}_{\!\varphi_{\!i}}\!(T_1, X_1, Y_1, T_2, X_2, Y_2)\;\;  \leftarrow \hspace{-8mm}&\\
&~~~~~~~~~~~~~~~~~~~~~~~~~~~{\it Match}_{\!\varphi_{\!i}}\!(T_2, X_2, Y_2, T_1, X_1, Y_1).~~~~(i \in \{1,2\})& \vspace{-8mm}
\end{align*}
}\vspace{-6mm}
{\small
\begin{align*}
&~{\it OldVersion}_{_R}\!(T_1,\bar{Z}_1)\;\;  \leftarrow \ R'(T_1,\bar{Z}_1), \; R'(T_1,\bar{Z}'_1),\hspace{-8mm}&\\
&~~~~~~~~~~~~~~~~~~~~~~~~~~~~~~~~~~~~~~~~~~~~~~~~~~~~~~~~~~~~\bar{Z}_1 \preceq \bar{Z}'_1, \; \bar{Z}_1 \neq \bar{Z}'_1.\hspace{-8mm}
\end{align*}
}
\vspace{-6mm}{\small
\begin{align*}
&~~~~~\;\;  \leftarrow \ {\it N\!otMatch}_{\!\varphi_{\!i}}\!( T_1, X_1, Y_1,T_2, X_2, Y_2),\hspace{25cm}&\\
&~~~~~~~~~~~~~~~~~~\hspace{15mm}{\it not} \; {\it Old\!Version}_{_R}\!(T_1, X_1, Y_1),\hspace{-18mm}&\\
&~~~~~~~~~~~~~~~~{\it not} \; {\it Old\!Version}_{_R}\!(T_2, X_2, Y_2).
~~~~~~~~~~~~~~~~~~ (i \in \{1,2\}) \hspace{-8mm}
\end{align*}}
\vspace{-6mm}{\small
\begin{align*}
&\hspace{-0.3cm}3.~R'(T_1, X_1, Y_3)\;\;  \leftarrow \ {\it Match}_{\!\varphi_{\!1}}\!(T_1,X_1, Y_1, T_2, X_2, Y_2),\hspace{-8mm}&\\
&~~~~~~~~~~~~~~~~~~~~~~~~~~~~~~~~~~~~~~~~~~~~~~~~~~~~~~~~~~~~~~~~~~M_{\!B}\!(Y_1, Y_2,Y_3).\hspace{-8mm}&\\
&R'(T_1, X_1, Y_3)\;\;  \leftarrow \  {\it Match}_{\!\varphi_{\!2}}\!(T_1,X_1, Y_1, T_2, X_2, Y_2),\hspace{-8mm}&\\
&~~~~~~~~~~~~~~~~~~~~~~~~~~~~~~~~~~~~~~~~~~~~~~~~~~~~~~~~~~~~~~~~~M_{\!B}\!(Y_1, Y_2,Y_3).\hspace{-8mm}
\end{align*}
}
\vspace{-6mm}
{\small
\begin{align*}
&~~~~4.~{\it Prec}(T_1, X_1, Y_1, T_2, X_2, Y_2, T_1, X_1, Y'_1, T_3, X_3,
Y_3)\;\;  \leftarrow \hspace{5mm}&\\
&~~~~~~~~~~~~~~~~~{\it Match}_{\!\varphi_{\!i}}\!(T_1, X_1, Y_1, T_2, X_2,
Y_2), \hspace{-8mm}&\\
&~~~~~~~~~~~~~~~~~~~~~~~~~~~{\it Match}_{\!\varphi_{\!j}}\!(T_1, X_1, Y'_1,
T_3, X_3, Y_3),\hspace{-8mm}&\\
&\hspace*{2.5cm}Y_1 \preceq Y'_1, \; Y_1 \neq Y'_1.~~~~~~~~~~~~~~~~~ (i,j \in \{1,2\})&
\end{align*}
} \vspace{-6mm}
{\small
\begin{align*}
&~~~~5.~{\it Prec}(T_1, X_1, Y_1, T_2, X_2, Y_2, T_1, X_1, Y_1, T_3, X_3,
Y_3)\;\;  \leftarrow \hspace{5mm}&\\
&~~~~~~~~~~~~\hspace{2.4cm}{\it Match}_{\!\varphi_{\!i}}\!(T_1, X_1, Y_1, T_2,
X_2, Y_2),\hspace{-8mm}&\\
&~~~~~~~~~~~~~~{\it Match}_{\!\varphi_{\!j}}\!(T_1,
X_1, Y_1, T_3, X_3, Y_3), \ M_{\!B}\!(Y_1,Y_3,Y_4),&\\
&~~~~~~~~~~~~~~~~~~~~~~~ \hspace{1cm}Y_1 \neq Y_4.~~~~~~~~~~~~~~~~~~~~~~~~~~~~~~~(i,j \in \{1,2\})&
\end{align*}}
\vspace{-6mm}
{\small
\begin{align*}
&\!\!\!\!6.~{\it Prec}(T_1, \bar{Z}_1, T_2, \bar{Z}_2,  T_1, \bar{Z}_1, T_2, \bar{Z}_2)\;\;  \leftarrow \hspace{10mm}&\\
&~~~~~~~~~~~~~~~~~~~~~~~~~~{\it Match}_{\!\varphi_{\!i}}\!(T_1, \bar{Z}_1, T_2, \bar{Z}_2). ~~~~~~~~~~~~~ (i \in \{1,2\})\hspace{-8mm}
\end{align*}
}

\vspace{-9mm}
{\small
\begin{align*}
&\;\;  \leftarrow  {\it Prec}(T_1, \bar{Z}_1, T_2, \bar{Z}_2,  T_1, \bar{Z}'_1, T_3, \bar{Z}_3),\hspace{5mm}&\\
&~~~~~~~~~~~~~~~~~~{\it Prec}(T_1, \bar{Z}'_1, T_3, \bar{Z}_3, T_1, \bar{Z}_1, T_2, \bar{Z}_2),\hspace{-8mm}&\\
&~~~~~~~~~~~~~~~~~~~~~~~~~~~~~~~~~~(T_1, \bar{Z}_1, T_2, \bar{Z}_2) \neq (T_1, \bar{Z}'_1, T_3, \bar{Z}_3). \hspace{-8mm}
\end{align*}
}

\vspace*{-10mm}
{\small
\begin{align*}
&\;\;  \leftarrow  {\it Prec}(T_1, \bar{Z}_1, T_2, \bar{Z}_2,  T_1, \bar{Z}'_1, T_3, \bar{Z}_3),\hspace{5mm}&\\
&~~~~~~~~~~~~~~~~~~{\it Prec}(T_1, \bar{Z}'_1, T_3, \bar{Z}_3, T_1, \bar{Z}''_1, T_4, \bar{Z}_4),\hspace{-8mm}&\\
&~~~~~~~~~~~~~~~~~~~~~~~~~~{\it not} \; {\it Prec}(T_1, \bar{Z}_1, T_2, \bar{Z}_2, T_1, \bar{Z}''_1, T_4, \bar{Z}_4). \hspace{-8mm}
\end{align*}
}
\vspace{-6mm}{\small
\begin{align*}
&\hspace{-0.15cm}7.~R^c(T_1, X_1, Y_1)\;\;  \leftarrow R'(T_1, X_1, Y_1),\hspace{5mm}&\\
&~~~~~~~~~~~~~~~~~~~~~~~~~~~~~~~~~~~~~~~~~~~~~~~~{\it not} \; {\it OldVersion}_{_R}\!(T_1, X_1, Y_1).\hspace{-8mm}
\end{align*}
}

\vspace{-6mm}  The program constraint under 2. (last in the list) ensures that all new, applicable matchings have to be eventually carried out. The last set of rules (one for each database predicate) collect the final, clean extensions of them.

Program $\Pi(D_0,\Sigma)$ has two stable models, whose $R^c$-atoms are shown below:

\vspace{1mm}
{\small \noindent $M_1 = \left\{...,R^c(t_1, a_1,b_{12}), R^c(t_2,a_{2}, b_{12}), R^c(t_3,a_3,b_3) \right\}$,

\vspace{1mm}

\noindent $M_2 = \{...,R^c(t_1,a_1,b_{123}), R^c(t_2,a_{2}, b_{123}), R^c(t_3,a_{3},b_{23})\}$.}

\vspace{1mm}
\noindent From them we can read off the two clean instances $D_1$, $D'_2$ for $D_0$ that
were obtained from the chase.
\boxtheorem
\end{example}

\ignore{
Next we establish that, for an instance $D_0$ and a set $\Sigma$ of MDs,
the set $\nit{SM}(\Pi(D_0,\Sigma))$ of the stable models of $\Pi(D_0,\Sigma)$
is in one-to-one correspondence with the set of $(D_0,\Sigma)$-clean instances.}

The cleaning program $\Pi(D_0,\Sigma)$ allows us to reason in the presence of uncertainty as represented by the possibly multiple
clean instances. Actually, it holds that there is a one-to-one correspondence between $\mc{C}(D_0,\Sigma)$ and
the set $\nit{SM}(\Pi(D_0,\Sigma))$ of
stable models of $\Pi(D_0,\Sigma)$. Furthermore, the program $\Pi(D_0,\Sigma)$ without its program constraints
belongs to the class
 $\nit{Datalog}^{\!\!\vee,\!\!\n\!\!,s}$, the subclass of programs in $\nit{Datalog}^{\!\!\vee,\!\!\n}$
that have {\em stratified negation} \cite{eiter95}. As a consequence, its stable models can be computed bottom-up by propagating data upwards from the
underlying extensional database (that corresponds to the set of {\em facts} of the program), and making sure to minimize the selection of true atoms from the disjunctive heads.
Since the latter introduces a form of non-determinism, a program may have several stable models. If the program is non-disjunctive, i.e. belongs to the $\nit{Datalog}^{\!\!\n\!\!,s}$, it has a single stable model that can be computed in polynomial time in the size of the extensional database $D$. The program constraints in $\Pi(D_0,\Sigma)$ make it unstratified \cite{gelfondBook}. However, this is not a crucial problem because they act as a filter, eliminating
the models that make them true from the class of models  computed with the bottom-up approach.

\vspace{-2mm}
\section{Relational MDs}\label{sec:relMDs}

We now introduce a class of  MDs  that have found useful applications in blocking for learning a classifier for ER \cite{sum15}. They allow bringing additional relational knowledge into the conditions of the MDs. Before doing so,
notice that an explicit formulation of the MD in (\ref{eq:md2}) in classical predicate logic is:\footnote{Similarity symbols can be treated as regular, built-in, binary predicates, but
 the identity symbol, $\doteq$, would be non-classical.}
{\small \begin{eqnarray}
\varphi\!:  \ \ \forall t_1 t_2 \ \forall \bar{x}_1 \bar{x}_2&\!\!\!\!\!\!\!\!(&\!\!\!\!\!\!\!\!R_1(t_1,  \bar{x}_1) \wedge R_2(t_2,  \bar{x}_2) \  \wedge \nonumber \\
&&\!\!\!\!\!\!\!\!\bigwedge_j x_1^j \approx_j x_2^j \ \  \longrightarrow \ \
 y_1 \doteq y_2),\label{eq:md12145}
\end{eqnarray}}
\phantom{ooo}

 \vspace{-6mm} \noindent with $x_1^j, y_1 \in \bar{x}_1, \ x_2^j, y_2\in \bar{x}_2$.  The $t_i$ are  variables for  tuple IDs.  \ignore{In (\ref{eq:md12145}), $\approx_j$ is a binary similarity relation  on domain $\nit{Dom}_j$.}  $\nit{LHS}(\varphi)$ and $\nit{RHS}(\varphi)$ denote the sets of
atoms on the LHS and RHS of $\varphi$,
respectively. \ Atoms $R_1(t_1, \bar{x}_1)$ and $R_2(t_2, \bar{x}_2)$ contain all the variables in the MD; and  similarity  and identity atoms involve one  variable from each of $R_1, R_2$.

Now, {\em relational MDs} may have in their  LHSs, in addition to the two {\em leading atoms}, as $R_1, R_2$ in (\ref{eq:md12145}), additional database atoms, from more than one relation, that are used to give context to similarity atoms
 in the MD, and capture additional relational knowledge via additional conditions. Relational MDs extend ``classical"  MDs.

\begin{example} \label{ex:blockMD} With predicates {\small $\nit{Author(AID,Name,}$ $\nit{PTitle,}$ $\nit{ABlock)},$  $\nit{Paper(PID,PTitle,Venue,PBlock)}$} (with ID and block attributes), this MD, $\varphi$, is  relational:
{\small \begin{eqnarray}
\underline{\nit{Author}(t_1,x_1,y_1,\nit{bl}_1)} \ \ \wedge \nit{Paper}(t_3,y_1',z_1, \nit{bl}_4) \ \wedge \ y_1 \approx y_1' \ \wedge \hspace{4mm} \nonumber &&\\ \underline{\nit{Author}(t_2,x_2,y_2,\nit{bl}_2)} \ \ \wedge \  \nit{Paper}(t_4,y_2',z_2, \nit{bl}_4) \ \wedge \ y_2 \approx y_2' \ \wedge
 \nonumber  \hspace{4mm}&&\\ x_1 \approx x_2 \ \wedge \ y_1 \approx y_2    \longrightarrow \ \ \nit{bl}_1 \doteq \nit{bl}_2,~~~~~~&&\nonumber %\label{eq:blmd}
 \end{eqnarray}}
\hspace*{-2mm} with implicit quantifiers, and underlined leading atoms (they contain the identified variables on the RHS). It  contains   similarity comparisons involving  attribute values for both relations  \nit{Author} and \nit{Paper}.
It specifies that when the \nit{Author}-tuple similarities on the LHS hold, and their papers are similar to those in corresponding \nit{Paper}-tuples that are in the {\em same} block (an implicit similarity captured by the join variable $\nit{bl}_4$), then blocks $\nit{bl}_1, \nit{bl}_2$ have to be made identical. This blocking policy
uses relational knowledge (the relationships between \nit{Author} and \nit{Paper} tuples), plus the blocking decisions already made about \nit{Paper} tuples.
\boxtheorem
\end{example}

\ignore{
\begin{example} \label{ex:blockMD}  This is  a relational MD:
{\small \begin{eqnarray}
\hspace*{-2mm}\underline{\nit{Author}(t_1,x_1,y_1,\nit{bl}_1)} \ \ \wedge & \hspace{-15mm}\nit{Paper}(t_3,y_1,z_1, \nit{bl}_4) \ \wedge &  \nonumber \\\underline{\nit{Author}(t_2,x_2,y_2,\nit{bl}_2)} \ \ \wedge \ & \hspace{-15mm}\nit{Paper}(t_4,y_2,z_2, \nit{bl}_4) \ \wedge &
 \nonumber \\ & \hspace{-7mm} x_1 \approx_1 x_2 \wedge z_1 \approx_2 z_2 \  \rightarrow \  \nit{bl}_1 \doteq \nit{bl}_2,& \nonumber %\label{eq:blmd}
 \end{eqnarray}}
\hspace*{-2mm} with implicit quantifiers, and underlined leading atoms (they contain the identified variables on the RHS). It  contains   similarity comparisons involving  attribute values for both relations  \nit{Author} and \nit{Paper}.
It specifies that when the \nit{Author}-tuple similarities on the LHS hold, and the corresponding \nit{Paper}-tuples are in the same block (captured by the join variable $\nit{bl}_4$), then blocks $\nit{bl}_1, \nit{bl}_2$ have to be made identical. This blocking policy
uses relational knowledge (the relationships between \nit{Author} and \nit{Paper} tuples), plus the blocking decisions already made about \nit{Paper} tuples.
\boxtheorem
\end{example} }

\vspace{-2mm}
\section{Single-Clean-Instance Classes}\label{sec:good}

First we introduce some notation. For an MD $\varphi$, $\nit{ALHS}(\varphi)$ denotes the set of (non-tid) attributes (with predicates) appearing
in similarities in the
  LHS of $\varphi$ (including  equalities, implicit or not). Similarly, $\nit{ARHS}(\varphi)$ contains the attributes appearing {\em in identities} in the RHS. In Example \ref{ex:blockMD}: {\scriptsize $\nit{ALHS}(\varphi) = \{\nit{Author[Name]},$ $\nit{Author[PTitle]},$
  $\nit{Paper[PTitle]},$ $\nit{Paper[PBlock]}\}$, \
$\nit{ARHS}(\varphi) =$  $\{\nit{Author[ABlock]}\}$}.

 As shown in \cite{tocs}, for the classical case of {\em similarity-preserving} MDs (i.e. whose MFs
 satisfy $a \approx_A a'$ implies $a \approx \match_A(a',a'')$), the chase-procedure computes  a single clean instance in polynomial time in the size of
 the initial instance. The same holds for the classical case of
 {\em non-interacting} MDs.   Now, a set $\Sigma$ of possibly relational  MDs is  {\em non-interacting} if there are  no $\varphi_1, \varphi_2\!\in\!\Sigma$ (possibly the same), with  $\nit{ARHS}(\varphi_1) \cap \nit{ALHS}(\varphi_2) \neq \emptyset$.   Relational similarity-preserving MDs are trivially defined by
 using similarity preserving MFs.
Through simple changes in the proofs given in \cite{tocs} for classical similarity-preserving and non-interacting MDs, it is possible to prove  that, for both classes, for a given initial instance $D$, there is a  single resolved instance that can be computed in polynomial time
in the size of $D$. We say that these classes of MDs have the {\em single-clean instance} property, in short, {\em they are SCI}.

There is another class of  {\em combinations of relational MDs $\Sigma$ and initial instances $D$} that lead to a single {\em clean} instance:\footnote{More precisely, it is  duplicate-free wrt. the MDs, i.e. no additional enforcements thereof are possible}
That of {\em similarity-free attribute intersection} (SFAI) combinations $(\Sigma,D)$. 

%{\small $\{\nit{Author(AID,Name,}$ $\nit{PTitle,}$ $\nit{ABlock)},$  $\nit{Paper(PID,PTitle,Venue,PBlock)}$}

\begin{definition}\label{def:SFAInew}
 Let $\Sigma$ be a set of relational MDs and $D$ an instance. The combination $(\Sigma,D)$ has the   SFAI property (or is SFAI) if,  for every $\varphi_1, \varphi_2 \in \Sigma$ (which could be the same) and attribute $R[A] \in \nit{ARHS}(\varphi_1)$ $\cap$ $\nit{ALHS}(\varphi_2)$,  it holds: If $S_1, S_2 \subseteq D$ with $R(\bar{c}) \in  S_1 \cap S_2$, then  $\nit{LHS}(\varphi_1)$ is false in $S_1$ or $\nit{LHS}(\varphi_2)$ is false in $S_2$.\footnote{We informally say that $\varphi_1$ is not applicable in $S_1$, etc.}  \boxtheorem
\end{definition}
Non-interacting sets of MDs are trivially SFAI for every initial instance $D$.
 In general, different orders of MD enforcements  may result in different clean instances, because tuple similarities may be broken  during the chase with interacting MDs and non-similarity-preserving MFs, without reappearing again \cite{tocs}.
With SFAI combinations, two similar tuples, i.e. with similar attribute values, in the original instance $D$ -or becoming similar along a chase sequence- may have the similarities broken
in a chase sequence, but they will   reappear later on in the same and the other chase sequences. Thus, different orders of MD enforcements cannot lead in the end to different clean instances.

%\comlb{A simple good example?}

Contrary to the {\em syntactic} class of non-interacting (relational) MDs and the MF-dependant class of similarity-preserving MDs,  SFAI is a {\em semantic} class that depends on the initial instance (but not
on subsequent instances obtained through the chase). Checking the SFAI property for $(\Sigma,D)$ can be done by posing Boolean conjunctive queries (with similarity built-ins) to $D$; actually for each pair
$\varphi_1,\varphi_2$  in $\Sigma$, a query, $\mc{Q}_{\varphi_1,\varphi_2}^A$, if $A \in \nit{ARHS}(\varphi_1) \cap \nit{ALHS}(\varphi_2)$, and a query, $\mc{Q}_{\varphi_2,\varphi_1}^B$, if  $B \in \nit{ARHS}(\varphi_2) \cap \nit{ALHS}(\varphi_1)$.\footnote{E.g.  $R\left[B\right] \approx R\left[B\right] \rightarrow R\left[A\right] \doteq R\left[A\right], \ R\left[A\right] \approx R\left[A\right] \rightarrow R\left[B\right] \doteq R\left[B\right]$ give rise to two SFAI tests (two queries).}

\begin{example}\label{ex:first2} (ex. \ref{ex:first1} cont.) Consider the same classical MDs and MFs, but now with  $a_1\approx a_2$, $b_3\approx b_4$, and
 new instance:

\vspace{-2mm}
\begin{multicols}{2}
{\small \begin{center}
\begin{tabular}{c|c|c|}\hline
$R(D)$&$A$ & $B$ \\ \hline
$t_1$&$a_1$ & $b_1$ \\
$t_2$&$a_2$ & $b_2$ \\
$t_3$&$a_3$ & $b_3$ \\
$t_4$&$a_4$ & $b_4$ \\\cline{2-3}
\end{tabular}
\end{center}}

\noindent The MDs are interacting, and both {\em applicable} on $D$, i.e. their LHSs are \ true. \ We can \ check the \ SFAI \ property \ for \ the
\end{multicols}

\vspace{-3mm}\noindent  combination ($\Sigma,D)$ posing the following, implicitly existentially quantified, Boolean conjunctive queries to $D$:\footnote{For each of the intersections: $\nit{ARHS}(\varphi_1) \cap \nit{ALHS}(\varphi_2) = \{R[B]\}$, and $\nit{ARHS}(\varphi_2) \cap \nit{ALHS}(\varphi_2) = \{R[B]\}$.}\vspace{-2mm}
{\small \begin{eqnarray*}
 \mc{Q}_{\varphi_1, \varphi_2}^{R[B]}\!:&&R(t_1, x_1, y_1) \ \wedge \ R(t_2, x_2, y_2) \ \wedge \ x_1 \approx x_2 \ \wedge\\&&~~~~~~~~~R(t_3, x_3, y_3) \ \wedge \ y_2\approx y_3,\\
  \mc{Q}_{\varphi_2, \varphi_2}^{R[B]}\!:&&R(t_1, x_1, y_1) \ \wedge \ R(t_2, x_2, y_2) \ \wedge \ y_1 \approx y_2 \ \wedge\\&&~~~~~~~~~R(t_3, x_3, y_3)   \wedge \ y_2\approx y_3.
\end{eqnarray*}}
\phantom{oo}

\vspace{-8mm}
\begin{multicols}{2}
\noindent which take the value {\em false} in $D$. Then, $(\Sigma,D)$ is SFAI. This is consistent with the easily verifiable observation that, no matter

{\small \begin{center}
\begin{tabular}{c|c|c|}\hline
$R(D')$&$A$ & $B$ \\ \hline
$t_1$&$a_1$ & $b_{12}$ \\
$t_2$&$a_2$ & $b_{12}$ \\
$t_3$&$a_3$ & $b_{34}$ \\
$t_4$&$a_4$ & $b_{34}$ \\\cline{2-3}
\end{tabular}
\end{center}}
\end{multicols}
\vspace{-4mm}\noindent how the MDs are applied, a single clean instance, $D'$ above,  is
always achieved. \boxtheorem \end{example}

The example  shows that it is possible to decide  in polynomial time in the size of $D$ if a combination $(\Sigma,D)$ is SFAI: The number of queries does not depend on
 $D$, and they can be answered in polynomial time in data.
 Furthermore, it is possible to prove from the definition and the chase
that SFAI (sets of) MDs  are also SCI. However, in Section \ref{sec:datalog} we will indirectly show that  this holds, by presenting stratified Datalog programs that implicitly represent the chase procedure based on them.
The SCI property  follows also from this.

\section{Datalog Programs for SRI Classes}\label{sec:datalog}

The general ASPs for classical MDs can be easily changed to deal with relational MDs, by including in the rule bodies the new relational atoms as extra conditions.

It is possible to take a set of MDs of the three kinds introduced in Section \ref{sec:good}, generate an ASP for them of the general form of Section \ref{sec:prel}, and next, appealing to a general semantic property in common for those three classes,   automatically rewrite the program
into a stratified Datalog program.

The rewriting is based on the facts that: (a) We do not need rules or constraints for the \nit{Prec} predicate, because
 imposing a linear order of matchings is not needed; basically all MDs can be applied in parallel. \ (b) For the same reason, we do not need disjunctive heads, as each applicable MD can be applied without affecting the results obtained by the applications of the others. That is, we do not have to
  withhold any matchings (via the \nit{NotMatch} predicates). \ (c) Old versions of tuples can be used in
future MDs enforcements without any undesirable impact on the result.  \ignore{{\bf In particular, we do not need  program constraints to eliminate models (chase sequences) obtained by matchings with old versions of tuples.}}

In essence, the semantic property of the three classes, which can be expressed and used as a systematic rewriting mechanism of the general cleaning ASP,  is that: {\em  When confronted with match or not match, we can safely match; and the matchings do not need to be linearly ordered. Also, old versions of tuples can be still used for matchings}.
The general transformation is illustrated  by means of an example.

\ignore{\comlb{The examples of rewriting go here.}
\begin{center}
$\varphi_1: R\left[A\right] \approx R\left[A\right] \rightarrow R\left[B\right] \doteq R\left[B\right]$,\\
$\varphi_2: R\left[B\right] \approx R\left[B\right] \rightarrow R\left[B\right] \doteq R\left[B\right]$.
\end{center}

{\small \begin{center}
$M_B(b_1, b_2,b_{12})$,

 $M_B(b_2, b_3,b_{23})$,

   $M_B(b_1, b_{23},b_{123})$.

 $M_B(b_3, b_4,b_{34})$.

  \end{center}} }

 \begin{example}\label{ex:GeneralToSpecial} (ex. \ref{ex:first2} cont.)   The general cleaning program  for $\Sigma$ in Example \ref{ex:first1} depends on the initial instance only through the
 program facts. Then, the same program can be used in Example \ref{ex:first2}, but with the facts corresponding to $R(D_0)$ replaced by those corresponding to $R(D)$. Since  $(\Sigma,D)$ is SFAI,
 the cleaning program can be automatically rewritten into the following residual program
 (with enumeration as Example \ref{ex:first1}):

\vspace{-1mm}
{\small \begin{enumerate}
\item
\ignore{$\nit{Dom}(a_i)$. ~~~~~~~~~~~$\nit{Dom}(b_i)$.~~~~~~~~~~~~($1 \leq i \leq 3$)\\
~$\nit{Dom}(b_{12})$. ~~~~~~~~~$\nit{Dom}(b_{23})$. ~~~~~~~~~$\nit{Dom}(b_{123})$.\\}
 $R(t_1, a_1, b_1).$  $R(t_2, a_2, b_2).$  $R(t_3, a_3, b_3).$    $R(t_4, a_4, b_4)$.
\ignore{$M_{\!B}\!(b_1, b_2,b_{12}).$ \ $M_{\!B}\!(b_2, b_3,b_{23}).$ \ $M_{\!B}\!(b_1, b_{23},b_{123}).$}
\end{enumerate} \vspace{-3mm}
\begin{eqnarray*}
~~~2.~{\it Match}_{\!\varphi_{\!1}}\!(T_1, X_1, Y_1,  T_2, X_2, Y_2)\; \; \leftarrow R(T_1, X_1, Y_1), \hspace{1.4cm}\\
R(T_2, X_2, Y_2),\;X_1 \approx X_2, \;  Y_1 \neq Y_2.\hspace{0.5cm}\\
{\it Match}_{\!\varphi_{\!2}}\!(T_1, X_1, Y_1, T_2, X_2, Y_2)\; \; \leftarrow R(T_1, X_1, Y_1), \hspace{1.4cm}\\
R(T_2, X_2, Y_2),\;Y_1 \approx Y_2, \;  Y_1 \neq Y_2.\hspace{0.5cm}\\
%\end{eqnarray*}\vspace{-3mm}
%\begin{eqnarray*}
{\it OldVersion}(T_1, X_1, Y_1)\;\;  \leftarrow \ R(T_1,X_1, Y_1), \hspace{2.2cm}\\  \hspace{-3.5cm} \; R(T_1, X_1, Y'_1), \ Y_1 \preceq Y'_1, \; Y_1 \neq Y'_1.\hspace{6mm}
\end{eqnarray*} \vspace{-7mm}
\begin{eqnarray*}
~~~3.~R(T_1, X_1, Y_3)\;\;  \leftarrow \ {\it Match}_{\!\varphi_{\!1}}\!(T_1, X_1, Y_1, T_2, X_2, Y_2),\hspace{3.2cm}\\   M_{\!B}\!(Y_1, Y_2,Y_3).\hspace{2.6cm}\\
R(T_1, X_1, Y_3)\;\;  \leftarrow \  {\it Match}_{\!\varphi_{\!2}}\!(T_1,X_1, Y_1, T_2, X_2, Y_2),\hspace{3.2cm}\\  M_{\!B}\!(Y_1, Y_2, Y_3).\hspace{2.6cm}\\ \vspace{-2mm}
%\end{eqnarray*}
%\begin{align*}
~~~7.~~R^c(T_1, X_1, Y_1)\; \leftarrow R(T_1, X_1, Y_1), \hspace{5.5cm}\\ {\it not} \; {\it OldVersion}(T_1, X_1, Y_1).\hspace{2.4cm}
%\end{align*}
\end{eqnarray*}
 }
 \phantom{ooo}

\vspace{-5mm}This program does not have disjunctive heads or program constraints. We still need the \nit{OldVersion} predicate to collect (the final versions of) the tuples in a clean instance. \boxtheorem
\end{example}

The general ASP programs of Section \ref{sec:prel} can be run on ASP solvers, such as {\em DLV} \cite{dlv,kr}. However, the specialized stratified Datalog programs of this section can be run with implementations of Datalog.
Actually, for their use in classification-based  ER reported in \cite{sum15}, the programs were specified using {\em LogicQL} and run on top of the Datalog-supporting {\em LogicBlox} platform \cite{logicblox}.

\section{Conclusions}

Matching dependencies (MDs) are an important addition to the declarative approaches to data cleaning, in particular, to the common and difficult problem of entity resolution (ER). We have shown that
MDs can be extended to capture additional semantic knowledge, which is important in applications, in particular, to machine learning.

Computing with MDs has a relatively high data complexity \cite{tocs}, but   some classes of MDs (possibly in combination with an instance) can be identified for which ER can be done in polynomial time
in data. Even more, it is possible to automatically produce Datalog programs that can be used to do ER with them.

%vspace{1mm}\noindent
%{\small {\bf Acknowledgements:} \ Research funded by NSERC Discovery.}

{\small
%{\scriptsize

}

\end{document}